\PassOptionsToPackage{table,dvipsnames}{xcolor}

\documentclass[12pt]{iopart}

\usepackage[svgnames]{xcolor}      % For colors
\usepackage{times}
\usepackage{geometry}
\usepackage[utf8]{inputenc}
\usepackage{enumitem,amssymb}
\usepackage{ragged2e}
\usepackage{graphicx}
\usepackage{comment}
\usepackage{multicol}
\usepackage[caption=false]{subfig}             
\definecolor{xlinkcolor}{cmyk}{1,1,0,0}
\usepackage{url}

\usepackage{titlesec}              % For custom sectioning
\usepackage{lmodern}  
\usepackage[numbers,square,sort&compress]{natbib}
\usepackage[
colorlinks=true,    % false: boxed links; true: colored links 
linkcolor=xlinkcolor,     % color of internal links            
citecolor=xlinkcolor,     % color of links to bibliography
filecolor=xlinkcolor,  % color of file links 
urlcolor=xlinkcolor,      % color of external link
final=true
]{hyperref}

\emergencystretch=\maxdimen
\usepackage{enumitem}
\setenumerate{itemsep=0mm}

\begin{document}

\newgeometry{margin=0in}  % temporarily remove margins
\thispagestyle{empty}     % remove page number

\newgeometry{left=0cm,right=0cm,top=0cm,bottom=0cm}  % Remove all margins
\thispagestyle{empty}  % Remove headers/footers and page number

\begin{figure}[t]
  \centering
  \includegraphics[width=0.999\paperwidth,height=0.999\paperheight]{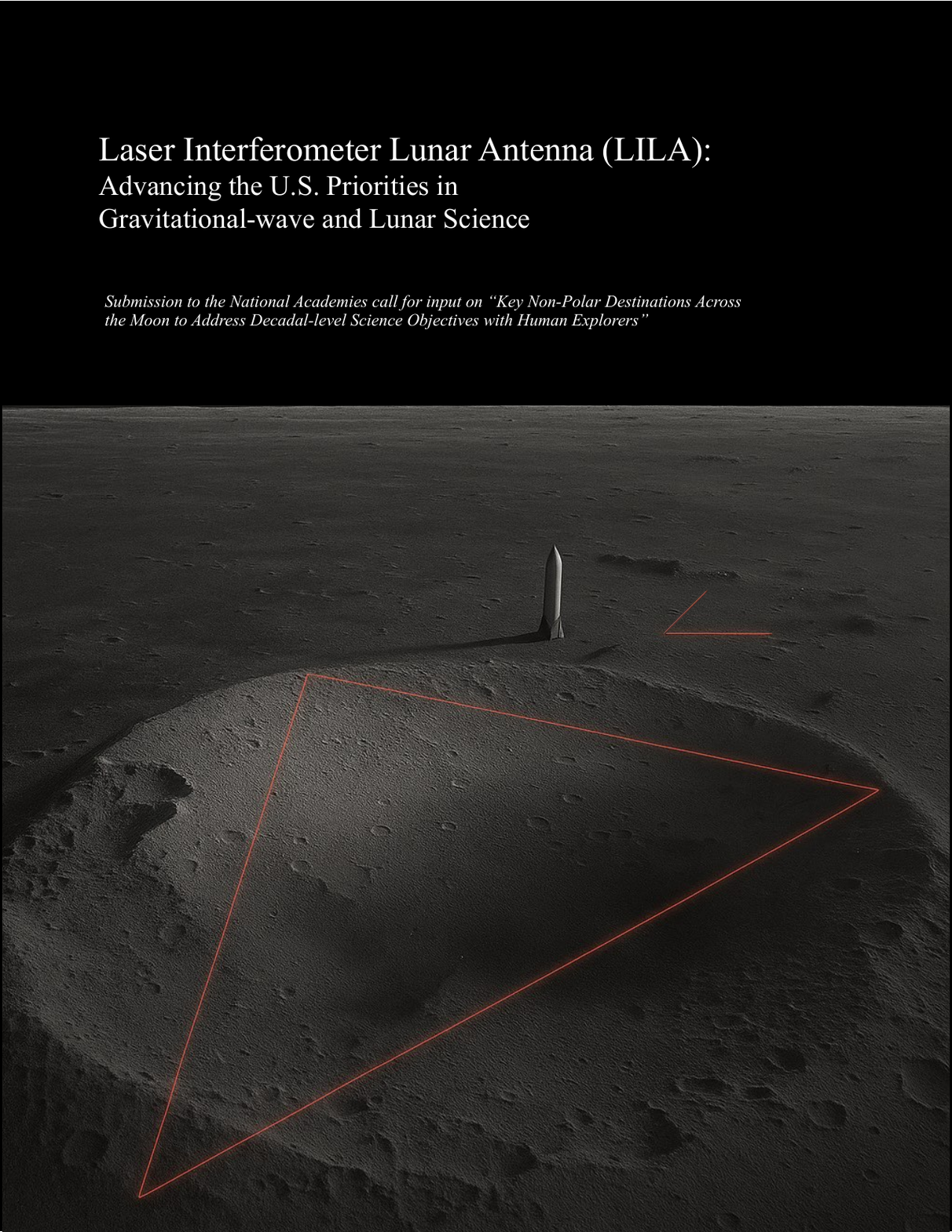}
\end{figure}

\clearpage
\restoregeometry  % Reset to your original margin

\title[Laser Interferometer Lunar Antenna]{Laser Interferometer Lunar Antenna (LILA): Advancing the U.S. Priorities in Gravitational-Wave and Lunar Science}

\author{
Karan Jani$^1$,
Matthew Abernathy$^2$,
Emanuele Berti$^3$,
Valerio Boschi$^4$,
Sukanya Chakrabarti$^5$,
Alice Cocoros$^6$,
John W. Conklin$^7$,
Teviet Creighton$^8$,
Simone Dell’Agnello$^9$,
Jean-Claude Diels$^{10}$,
Stephen Eikenberry$^{11}$,
T. Marshall Eubanks$^{12}$,
Kiranjyot Gill$^{13}$,
Jonathan E. Grindlay$^{13}$,
Kris Izquierdo$^6$,
Jaesung Lee$^{11}$,
Abraham Loeb$^{14}$,
Philippe Lognonné$^{15}$,
Francesco Longo$^{16}$,
Manuel Pichardo Marcano$^{17}$,
Mark Panning$^{18}$,
Paula do Vale Pereira$^{11}$,
Volker Quetschke$^{19}$,
Ashique Rahman$^6$,
Massimiliano Razzano$^4$,
Robert Reed$^{20}$,
Brett Shapiro$^6$,
David Shoemaker$^{21}$,
William Smith$^1$,
James Trippe$^{20}$,
Eric Van Stryland$^{11}$,
Wan Wu$^{22}$,
Anjali B. Yelikar$^1$
}

\address{$^1$Vanderbilt University, Vanderbilt Lunar Labs Initiative and Department of Physics and Astronomy, Nashville, TN 37235, USA}
\address{$^2$Air and Missile Defense Sector, The Johns Hopkins University Applied Physics Lab, Laurel, MD 20723, USA}
\address{$^3$Johns Hopkins University, William H. Miller III Department of Physics and Astronomy, 3400 N. Charles Street, Baltimore, MD 21218, USA}
\address{$^4$INFN, Largo Pontecorvo 3, I-56127 Pisa, Italy}
\address{$^5$University of Alabama, Huntsville, Department of Physics and Astronomy, 301 Sparkman Drive, Huntsville AL 35899, USA}
\address{$^6$The Johns Hopkins University Applied Physics Laboratory, Space Exploration Sector, Laurel, MD 20723, USA}
\address{$^7$University of Florida, Department of Mechanical and Aerospace Engineering, Gainesville, FL 32611, USA}
\address{$^8$University of Texas Rio Grande Valley, Center for Advanced Radio Astronomy, Brownsville, TX 78521, USA}
\address{$^9$INFN - Frascati National Labs, V. Enrico Fermi 54, Frascati 00044 (RM), Italy}
\address{$^{10}$University of New Mexico, Center for High Technology Materials and Department of Physics and Astronomy, 1313 Goddard St. SE, Albuquerque, NM 87106, USA}
\address{$^{11}$University of Central Florida, CREOL, The College of Optics and Photonics and Department of Physics, Orlando, FL 32816, USA}
\address{$^{12}$Space Initiatives Inc, Princeton, WV 24740, USA}
\address{$^{13}$AstroAI; Center for Astrophysics \textbar{} Harvard \& Smithsonian, 60 Garden Street, Cambridge, MA 02138-1516, USA}
\address{$^{14}$Center for Astrophysics \textbar{} Harvard \& Smithsonian, 60 Garden Street, Cambridge, MA 02138-1516, USA}
\address{$^{15}$Université Paris Cité, Institut de physique du globe de Paris, CNRS, Paris, France}
\address{$^{16}$University of Trieste, Department of Physics, Via Valerio 2, I-34127 Trieste, Italy}
\address{$^{17}$Fisk University \& Vanderbilt University, Vanderbilt Lunar Labs Initiative and Department of Life and Physical Science, Nashville, TN 37235, USA}
\address{$^{18}$Jet Propulsion Laboratory, California Institute of Technology,  Pasadena, CA 91109, USA}
\address{$^{19}$University of Texas Rio Grande Valley, South Texas Space Science Institute, Brownsville, TX 78512, USA}
\address{$^{20}$Vanderbilt University, Vanderbilt Lunar Labs Initiative and Department of Electrical and Computer Engineering, Nashville, TN 37235, USA}
\address{$^{21}$Massachusetts Institute of Technology, MIT Kavli Institute, Cambridge, MA 02139, USA}
\address{$^{22}$NASA Langley Research Center, Science Directorate, Hampton, VA 23681, USA}

\ead{karan.jani@vanderbilt.edu}

% P.Lognonné, Université Paris Cité, Institut de physique du globe de Paris, CNRS, Paris, France. 

%\clearpage

\begin{abstract}
    The Laser Interferometer Lunar Antenna (LILA) is a next-generation gravitational-wave (GW) facility on the Moon. By harnessing the Moon's unique environment, LILA fills a critical observational gap in the mid-band GW spectrum ($0.1 - 10$~Hz) between terrestrial detectors (LIGO, Virgo, KAGRA) and the future space mission LISA. Observations enabled by LILA will fundamentally transform multi-messenger astrophysics and GW probes of fundamental physics.
    %LILA observations will transform multi-messenger astrophysics and GW probes of fundamental physics. 
    LILA will measure the lunar deep interior better than any existing planetary seismic instruments.
    The LILA mission is designed for phased development aligned with capabilities of the U.S.’s Commercial Lunar Payload Services and Artemis programs. LILA is a unique collaboration between universities, space industries, U.S. government laboratories, and international partners.  
    %LILA is designed for phased deployment aligned with U.S. lunar exploration goals, with collaboration from industry and international partners.
\end{abstract}

%\section*{\task{Final draft edit: David S., Robert R.}} 
%\noindent

%   The Laser Interferometer Lunar Antenna (LILA) is a next-generation gravitational-wave (GW) facility on the surface of the Moon. By taking advantage of Moon’s unique environment, LILA will fill a critical observational gap in the mid-band GW spectrum ($0.1 - 10$~Hz) between terrestrial detectors such as LIGO, Virgo and KAGRA and the European Space Agency’s future mission LISA. Observations enabled by LILA will fundamentally transform multi-messenger astrophysics and GW probes of fundamental physics. LILA will also measure the deep lunar seismic activity more completely than any existing planetary seismic instrument. The LILA mission will be done in phases aligned with the capabilities of the U.S.’s Commercial Lunar Exploration Program and developed by a collaboration between universities, the space industry, U.S. government laboratories, and international partners. 

%LILA is designed to leverage the unique lunar environment to explore previously inaccessible regions of the GW and seismic lunar spectrums. In particular, LILA will fill a critical observational gap between terrestrial detectors such as LIGO and the ESA's space-based LISA mission, revolutionizing our understanding of the universe through groundbreaking multi-messenger astrophysics and of the lunar deep interior through ultra-sensitive seismic monitoring .

\section{Science Goals and Decadal-Level Objectives}
%\task{Section edits: Karan, TC: We do have to be careful in what we promise; LILA can reach sensitivity comparable to Big Bang Observer in terms of raw $S_h(f)$ but we still have only one detector.  Added a bullet for your consideration. Jasmine/Bill - revised this section to focus on only four bullets for condensing.}

%A NASA-appointed study in the late 1980s~\cite{Stebbins1990}, followed by recent work from recent international groups~\cite{GLOC, Harms_2021, LION, Stavros, Branchesi2023, Lunarmodes}, demonstrated that a lunar detector can uniquely access the deci-hertz gravitational-wave spectrum. LILA’s objectives thus significantly advance astronomy, fundamental physics, and lunar science.

\begin{figure}[ht!]
    \centering
    \includegraphics[trim = 0 0 0 0, clip, width = 0.96\textwidth]{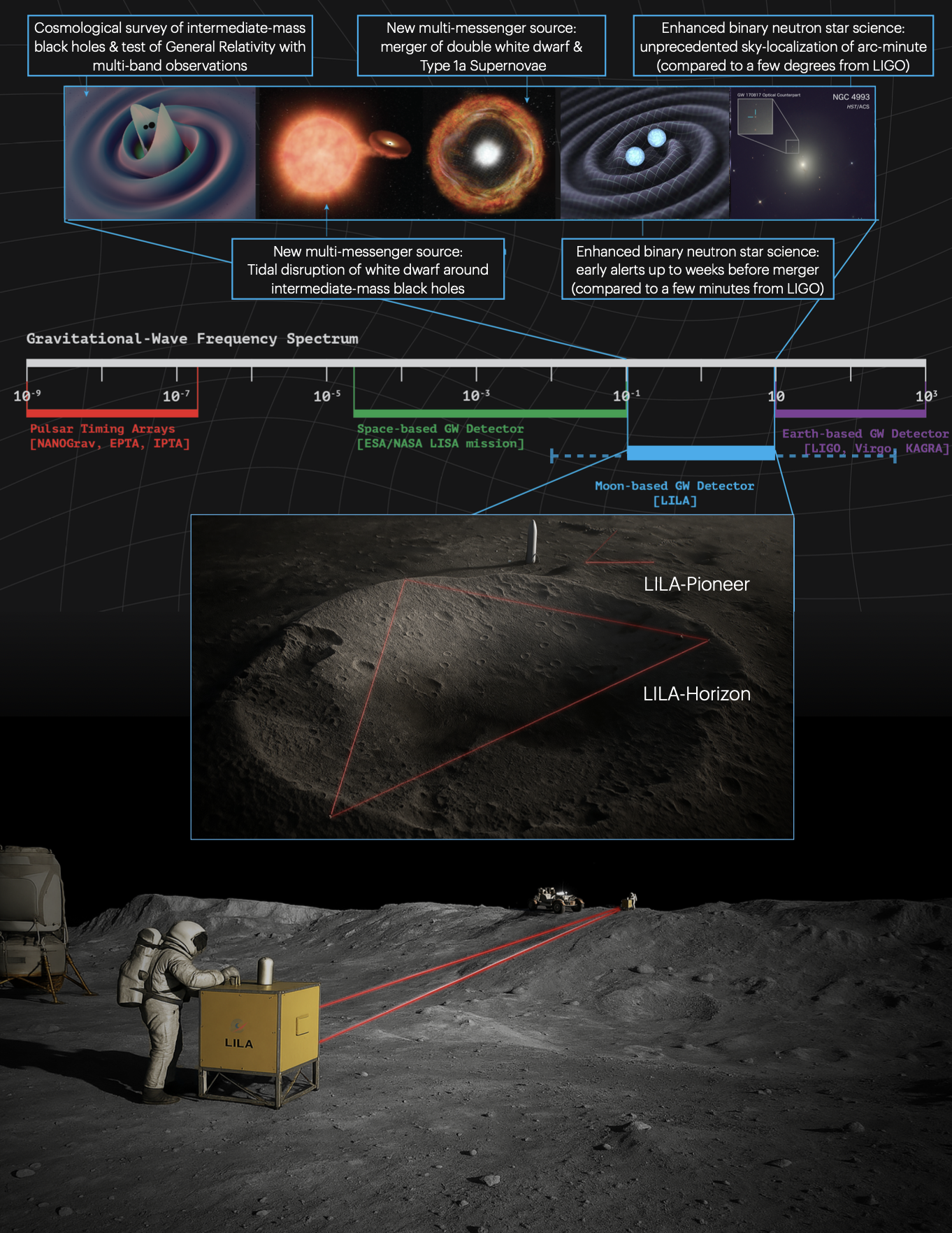}
    \caption{\textit{Top: }The entire GW detection spectrum. LILA science targets are highlighted. \textit{Middle:} Phases of the LILA Mission. \textit{Bottom: }Astronaut deploying LILA-Pioneer.}
    \label{figure:LILA_poster}
\end{figure}

The LILA mission supports the top priorities in Pathways to Discovery in Astronomy and Astrophysics for the 2020s~\cite{NAP26141}, Origins, Planetary Science, and Astrobiology Decadal 2023-2032~\cite{NAP26522}, and Biological and Physical Sciences Decadal 2023-2032~\cite{NAP26750}, and its development is aligned with NASA's Moon-to-Mars Objective: ``Advance understanding of physical systems and fundamental physics by utilizing the unique environments of the Moon." Observations enabled by LILA will advance astronomy, fundamental physics, and lunar science (Fig. \ref{figure:LILA_poster}): 

\begin{itemize}
    
    \item \textbf{New Windows on the Dynamic Universe:} 
    LILA will provide days to months notice for stellar binary mergers of neutron stars and black holes, with $\sim 1~\mathrm{arcmin}^2$ localization~\cite{GLOC}. These capabilities will revolutionize electromagnetic follow-up and enable precession measurements of the neutron star equation of state and gamma-ray burst jet structure~\cite{Birnholtz:2013bea, Yu:2020drq}. LILA will routinely detect multi-messenger sources such as Type Ia supernovae progenitors~\cite{PhysRevLett.106.201103,marcano2025}, core-collapse supernovae~\cite{gill2024}, and tidal disruption around intermediate-mass black holes (IMBHs)~\cite{Mandel_2018}. LILA's multi-messenger observations will enable a wide range of ``bright and dark sirens'' of cosmic expansion, providing an independent, sub-percent precision constraint on cosmological parameters (Hubble constant, dark energy equation of state, and dark matter density). 
%    These capabilities will revolutionize electromagnetic follow-up, uncover dark sirens at high redshift
    %These capabilities will revolutionize electromagnetic follow-up, uncover dark sirens at high redshift, and enable precision measurements of Type Ia supernova progenitors and gamma-ray burst jet structure~\cite{Birnholtz:2013bea, Yu:2020drq, PhysRevLett.106.201103}.} 
    
    \item \textbf{Unveiling the Drivers of Galaxy Growth:} 
    %LILA will measure cosmic acceleration using five classes of standard sirens~\citep{Mandel_2018}, allowing sub-percent precision constraints of parameters like the Hubble constant, dark energy equation of state, and dark matter density. 
    LILA will conduct a cosmological survey of IMBHs across their entire mass-range from $10^2\sim10^6~M_\odot$~\cite{Greene_2020}, providing the most definite understanding of their link to the growth of supermassive black holes~\cite{Natarajan_2020,Gair:2009gr, Pacucci_2020}. LILA's precise measurement of the masses, spins, and environments of IMBHs will be critical to probe the formation channels of these elusive black holes in the early universe.

%    intermediate-mass black hole (IMBH) mergers, including those in galactic centers, enabling insight into black hole formation, growth, and co-evolution with galaxies~\cite{Gair:2009gr, Pacucci_2020}. 
    
    \item \textbf{Search for New Physics:} 
    LILA will test General Relativity in previously unexplored regimes through multi-band measurements of black holes with LIGO and LISA~\cite{Jani_nature}. Such observations may reveal signatures of alternative gravity theories and imprints of dark matter, such as axion clouds, around post-merger black holes~\citep{sedda2019missing}. LILA will probe sub-solar mass binaries to 30\% of the observable universe, enabling new tests of dark matter candidates~\cite{Shandera_2018, 2006ApJ...653L..53A}. LILA offers an opportunity to measure the cosmological stochastic GW background from the Big Bang~\cite{Harry:2006}.  
    
%    LILA-Horizon can achieve sensitivities comparable to the Big Bang Observer for detecting early universe GWs~\cite{Harry:2006}. 
    %LILA will enable the first detection of GWs from core-collapse supernovae, revealing the explosion mechanism, progenitor structure, and stellar remnant fate~\cite{gill2024}. It will also detect GWs from extreme stellar collapse, including pair-instability supernovae, expanding understanding of explosion diversity~\cite{2011ApJ...730...70O, 2017ApJ...836..244W}. 
    
    \item \textbf{Deep Interior Lunar Geophysics}  
    LILA will resolve the Moon's normal modes for the first time, including inner core modes and those sensitive to deep interior lateral variations ~\cite{doi:10.1098/rsta.2023.0066}. LILA will enable 3D reconstruction of the Moon's internal structure, including inner core shape, core-mantle boundaries, and thermal gradients~\cite{Lognonne2024}, advancing top geophysical priorities by constraining the Theia–Earth impact geometry, differentiation, and mantle dynamics.

%LILA will resolve the Moon’s normal modes for the first time, including the inner core modes and those sensitive to deep interior lateral variations. These measurements will enable 3D reconstruction of the Moon’s internal structure, including the shape of the inner core, core-mantle boundaries, and thermal gradients~\cite{Lognonne2024}. This will directly advance OWL’s top geophysical priorities by constraining the Theia–Earth impact geometry, differentiation, and mantle dynamics.}

\end{itemize}

\section{Phase Implementation for LILA Mission}%~\task{Full section edit: Volker}}

%\task{Karan's note: discuss more on the technology, the payloads, the lander capability to have them - remove science or human role since other sections covers it}

Based on Apollo data~\citep{NASA1977, lognonne2009, PhysRevD.111.044061}, studies have shown that Moon's low seismic noise opens the mid-band GW spectrum ~\cite{GLOC, Stebbins1990, StebbinsDrever1990, Izumi_2021, Harms_2021,  Stavros, Branchesi2023, Lunarmodes}.   
%Lunar seismic noise in the GW mid-band is significantly lower than on Earth~\cite{PhysRevD.111.044061, Izumi_2021}. 
The surface lacks atmospheric disturbances (wind, pressure variations, humidity) and lacks anthropogenic or acoustic noise sources. The natural vacuum surpasses LIGO's ultra-high vacuum. Therefore, building LILA requires no major lunar infrastructure. Unlike space missions like LISA that have a limited lifetime due to orbital pull, LILA will operate for a decade or longer. The deployment of LILA is structured into two phases: 
%Our Moon is the only known location in the Solar System where sub-Hz interferometry is possible with current or near-term technologies (see the footnote below).
%The deployment of LILA is structured into two complementary phases:

\begin{itemize}
    \item \textbf{LILA-Pioneer (near term):} A baseline interferometer (3 to 5 km) deployed with the current generation of CLPS lander (Fig. \ref{figure:LILA_poster}-bottom). It will use lunar normal modes and optical readout to conduct the first mid-band GW survey \cite{Creighton_et_al_inprep}. The primary payload hosts the optical setup, seismometer, camera, and on-board computing, all inside a dust-protected, thermally managed case atop scaffolding. The secondary payloads consists of a retroreflector and seismic monitor.

    \item \textbf{LILA-Horizon (long term):} %\task{Section edits: Brett, Stephen}} 
    A 40-km three-station interferometer requiring larger cargo capacity and expansive surface mobility. Each station hosts a seismic isolation system (suspended test mass, anti-springs, active seismic deco-relation) and a quantum-enhanced laser frequency comb sensor, GravComb~\cite{10.1117/12.3047830}. LILA-Horizon delivers two transformative science returns: (i) milli-hertz to kilo-hertz broadband GW sensitivity, enabling multi-band, multi-messenger astrophysics from a single lunar site; (ii) massive improvement in mid-band GW sensitivity over LILA-Pioneer, enabling detection to the cosmic horizon.

\end{itemize}

\section{Site-Agnostic Deployment and Location Requirements} %\task{Section edits: James}}

LILA is viable on any lunar site; however, optimal scientific returns and minimal engineering complexity require:

%LILA can perform its science mission on any lunar site. However, careful site selection will optimize scientific return and minimize engineering complexity. Considerations are:

\begin{itemize}
\item \textbf{Line of sight:} 
End stations have to placed at few $\sim$meters elevation from primary station fir LILA-Pioneer ($\sim$km for LILA-Horizon) for an unobstructed line of sight.

\item  \textbf{Lunar Terrain Vehicle:} LILA-Pioneer must conform to range and terrain constraints for deployment of end stations. 

\item \textbf{Shallow moonquakes:} 
These may occur near lunar scarps~\cite{WATTERS2010}, which should be avoided due to rare landslides~\cite{Watters2025}. 

\item \textbf{Lunar environment:} 
Dust, radiation, and thermal factors favor non-equatorial sites~\cite{Horányi2015, DEANGELIS2007169, WILLIAMS2017300}.

\item \textbf{Anthropogenic noise:} Regions near other human activities (such as Artemis Base Camp at South Pole) should be avoided~\cite{Glanzer_2023}.

\item \textbf{Co-location of detectors:} LILA-Horizon should be established near LILA-Pioneer to ensure characterization of lunar environment on GW sensitivity.  %, and co-location ensures only no duplication of astronaut presence. 
\end{itemize}
%The site selection and deployment for LILA+ will be based on the performance and environmental characterization obtained from precursor LILA. Therefore, it is strategic to find a LILA site that accounts for the additional baseline requirement and buffer zone for LILA+.   

\section{Role of Human Presence} %\task{Section edits: Mark, James}}

%\task{Karan's note: Based on Artemis IV what are the things we can ask human for intial LILA? Have to make very specific request for humans - not simply its chepaer or less complex. Mark's note: I think this is fine, but we need to walk a fine line between emphasizing we need astronauts, and making it sound like we're going to need too much time from the astronauts. Best case is to emphasize things humans can do relatively quickly but are hard robotically, like initial alignment/leveling. I worry about the sections on mitigation and trouble shooting making it sound like this will be a constant astronaut time sink to get anything at all. Maybe combine those into one topic like "Ongoing Maintenance (as needed) Be more specific"}

%Occasional human presence at LILA's lunar site is required for success. 
%Occasional human presence at both LILA sites will be required for success of the science mission. Activities include:
Although LILA is designed for autonomous operations, the presence of astronauts will be critical for:
\begin{itemize}
    \item \textbf{Modular Assembly and Alignment:} Astronauts will perform the  LILA-Pioneer deployment and alignment; tasks difficult to achieve robotically. LILA-Horizon will be designed to be modular for easy assembly within a few days. 

    \item \textbf{Maintenance (as needed):} 
    Minimal ``hands-on" maintenance will be needed ($\sim$every five years) for occasional dust cleaning from optics or actuators, re-coating baffles, and manual retuning after modular replacements. Astronauts, guided by on-Earth specialists, can perform upgrades for LILA-Horizon.

%    Astronauts can do an efficient job of routine maintenance (eg: removing dust from lenses, repainting matte-black for stray light prevention surfaces); %LILA-Horizon may occasionally also need technicians to perform measurements, alignments, and repairs to maintain performance. 

 %   \item \textbf{Module replacement:} Astronauts will replace LILA modules as needed, approximately every five years, due to the Moon's harsh environment, extending LILA's lifetime into decades. 

%    \item \textbf{Routine maintenance:} Minimal ``hands-on" maintenance will be needed ($\sim$every five years) without loss of functionality. Routine maintenance includes occasional dust cleaning from optics or actuators, re-coating baffles, and manual retuning after modular replacements.
    
\end{itemize}

\section{Pre-placed Assets for Human Sortie} %\task{Section edits: Philippe, James}}

%\task{Karan's note: importance of FSS or LILA shoe box tech demo can also go here including other things. }

Before crewed deployment, robotic pre-positioning of support systems will maximize sortie efficiency:

\begin{itemize}

    \item \textbf{Seismic detectors} (e.g., a LILA tech demo) will assess local moonquake activity, helping avoid hazardous sites and calibrate baseline models.

    \item \textbf{Pre-placed equipment} will include tools, modular parts, auto-alignment systems, power units, environmental monitors, robotic assistants, and comms relays—allowing a 10-day astronaut sortie to complete deployment.

\end{itemize}

%\section{Strategic Alignment with Moon-to-Mars Objectives}

%LILA strategically aligns with NASA’s Moon-to-Mars objectives by leveraging the lunar environment to advance fundamental physics research. The experience gained from deploying and operating a sophisticated gravitational-wave observatory on the Moon will directly inform future exploration missions to Mars and beyond, demonstrating crucial technological and operational capabilities essential for sustainable human exploration.

\section{International collaboration}
Foreign partners will contribute to science and technology development of both LILA-Pioneer and LILA-Horizon. Potential hardware contributions include CLPS-ready Optical Very Broad Band seismometers and retroreflectors.
\\
\\
\\
\noindent(999/1000 words)
%\section{Conclusion}

%LILA will be capable of exploring fundamental GW physics, observing unique astronomical phenomena, and revealing the interior of the moon. LILA-Pioneer, utilizing only modern technology, will provide foundational science in the mHz regime and demonstrate capability. LILA-Horizon, employing more advanced technologies, will provide broadband capabilities beyond any current detector. This phased approach takes advantage of the planned U.S. lunar development and astronaut presence to deliver science addressing key decadal objectives.

%The Laser Interferometer Lunar Antenna exemplifies optimal utilization of lunar resources to achieve groundbreaking scientific objectives. By establishing an astrophysical observatory on the Moon’s non-polar farside, LILA will deliver unparalleled contributions to fundamental physics and astrophysics, while simultaneously exploring the deep lunar interior with unprecedent capabilities. Its phased deployment approach strategically aligns with ongoing and future Artemis missions, reinforcing the Moon’s role as a hub for sustained scientific discovery and exploration leadership.

%\clearpage

%\begin{figure*}[t!]
%\centering
% \includegraphics[scale=0.1,trim = {0 0 0 0}]{Figures/LILA_concept.png}
%\caption{{Astronaut deploying the LILA detector.} }
%\label{fig:LILA_site}
%\end{figure*}

%\clearpage

\pagestyle{plain}  % removes headers and footers
\markboth{}{}      % clears existing marks
\bibliography{references2}

\providecommand{\newblock}{}
\begin{thebibliography}{10}
\expandafter\ifx\csname url\endcsname\relax
  \def\url#1{{\tt #1}}\fi
\expandafter\ifx\csname urlprefix\endcsname\relax\def\urlprefix{URL }\fi
\providecommand{\eprint}[2][]{\url{#2}}

\bibitem{NAP26141}
{National Academies of Sciences, Engineering, and Medicine} 2023 {\em Pathways to Discovery in Astronomy and Astrophysics for the 2020s\/} (Washington, DC: The National Academies Press) ISBN 978-0-309-46734-6 \urlprefix\url{https://nap.nationalacademies.org/catalog/26141/pathways-to-discovery-in-astronomy-and-astrophysics-for-the-2020s}

\bibitem{NAP26522}
{National Academies of Sciences, Engineering, and Medicine} 2023 {\em Origins, Worlds, and Life: A Decadal Strategy for Planetary Science and Astrobiology 2023-2032\/} (Washington, DC: The National Academies Press) ISBN 978-0-309-47578-5 \urlprefix\url{https://nap.nationalacademies.org/catalog/26522/origins-worlds-and-life-a-decadal-strategy-for-planetary-science}

\bibitem{NAP26750}
{National Academies of Sciences, Engineering, and Medicine} 2023 {\em Thriving in Space: Ensuring the Future of Biological and Physical Sciences Research: A Decadal Survey for 2023-2032\/} (Washington, DC: The National Academies Press) ISBN 978-0-309-69498-8 \urlprefix\url{https://nap.nationalacademies.org/catalog/26750/thriving-in-space-ensuring-the-future-of-biological-and-physical}

\bibitem{GLOC}
Jani K and Loeb A 2021 {\em Journal of Cosmology and Astroparticle Physics\/} {\bf 2021} 044 ISSN 1475-7516 \urlprefix\url{http://dx.doi.org/10.1088/1475-7516/2021/06/044}

\bibitem{Birnholtz:2013bea}
Birnholtz O and Piran T 2013 {\em Phys. Rev. D\/} {\bf 87} 123007 (\textit{Preprint} \eprint{1302.5713}) \urlprefix\url{https://journals.aps.org/prd/abstract/10.1103/PhysRevD.87.123007}

\bibitem{Yu:2020drq}
Yu Y~W 2020 {\em Astrophys. J.\/} {\bf 897} 19 (\textit{Preprint} \eprint{2001.00205})

\bibitem{PhysRevLett.106.201103}
Falta D, Fisher R and Khanna G 2011 {\em Phys. Rev. Lett.\/} {\bf 106}(20) 201103 \urlprefix\url{https://link.aps.org/doi/10.1103/PhysRevLett.106.201103}

\bibitem{marcano2025}
{Pichardo Marcano} M, {Yelikar} A~B and {Jani} K 2025 {\em arXiv e-prints\/} arXiv:2503.04936 (\textit{Preprint} \eprint{2503.04936})

\bibitem{gill2024}
{Gill} K 2024 {\em arXiv e-prints\/} arXiv:2405.13211 (\textit{Preprint} \eprint{2405.13211})

\bibitem{Mandel_2018}
Mandel I, Sesana A and Vecchio A 2018 {\em Classical and Quantum Gravity\/} {\bf 35} 054004 ISSN 1361-6382 \urlprefix\url{http://dx.doi.org/10.1088/1361-6382/aaa7e0}

\bibitem{Greene_2020}
Greene J~E, Strader J and Ho L~C 2020 {\em Annual Review of Astronomy and Astrophysics\/} {\bf 58} 257--312

\bibitem{Natarajan_2020}
Natarajan P 2020 {\em Monthly Notices of the Royal Astronomical Society\/} {\bf 501} 1413–1425 ISSN 1365-2966 \urlprefix\url{http://dx.doi.org/10.1093/mnras/staa3724}

\bibitem{Gair:2009gr}
Gair J~R, Mandel I, Sesana A and Vecchio A 2009 {\em Class. Quant. Grav.\/} {\bf 26} 204009 (\textit{Preprint} \eprint{0907.3292})

\bibitem{Pacucci_2020}
Pacucci F and Loeb A 2020 {\em The Astrophysical Journal\/} {\bf 895} 95 ISSN 1538-4357 \urlprefix\url{http://dx.doi.org/10.3847/1538-4357/ab886e}

\bibitem{Jani_nature}
{Jani} K, {Shoemaker} D~M and {Cutler} C 2020 {\em Nature Astronomy\/} {\bf 4} 260--265 (\textit{Preprint} \eprint{1908.04985})

\bibitem{sedda2019missing}
Sedda M~A, Berry C~P~L, Jani K {\em et~al.\/} 2019 The missing link in gravitational-wave astronomy: Discoveries waiting in the decihertz range (\textit{Preprint} \eprint{1908.11375})

\bibitem{Shandera_2018}
Shandera S, Jeong D and Gebhardt H~S~G 2018 {\em Physical Review Letters\/} {\bf 120} ISSN 1079-7114 \urlprefix\url{http://dx.doi.org/10.1103/PhysRevLett.120.241102}

\bibitem{2006ApJ...653L..53A}
{Amaro-Seoane} P and {Freitag} M 2006 {\em Astrophysical Journal Letters\/} {\bf 653} L53--L56

\bibitem{Harry:2006}
{Harry} G~M, {Fritschel} P, {Shaddock} D~A, {Folkner} W and {Phinney} E~S 2006 {\em Classical and Quantum Gravity\/} {\bf 23} 4887--4894

\bibitem{doi:10.1098/rsta.2023.0066}
Cozzumbo A, Mestichelli B, Mirabile M, Paiella L, Tissino J and Harms J 2024 {\em Philosophical Transactions of the Royal Society A: Mathematical, Physical and Engineering Sciences\/} {\bf 382} 20230066 (\textit{Preprint} \eprint{https://royalsocietypublishing.org/doi/pdf/10.1098/rsta.2023.0066}) \urlprefix\url{https://royalsocietypublishing.org/doi/abs/10.1098/rsta.2023.0066}

\bibitem{Lognonne2024}
Lognonné P, Panning M, Tauzin B, Kawamura T and Jani K 2024 Deploying extreme sensitive laser strainmeter and distributed accoustic sensing on future artemis opportunities: geophysical goals and challenges. {Abstract No.\ 5042,} in \textit{Annual Meeting of the Lunar Exploration Analysis Group (LEAG)}, October 28--30, 2024, Houston, TX, USRA/LPI available at \url{https://www.hou.usra.edu/meetings/leag2024/pdf/5042.pdf}, accessed July 24, 2025

\bibitem{NASA1977}
Giganti J~J, Larson J, JP, Richard, Tobias R and Weber J 1977 Lunar surface gravimeter experiment Tech. Rep. NASA-TM-X-73093 National Aeronautics and Space Administration \urlprefix\url{https://ntrs.nasa.gov/api/citations/19770012037/downloads/19770012037.pdf}

\bibitem{lognonne2009}
Lognonné P, Le~Feuvre M, Johnson C~L and Weber R~C 2009 {\em Journal of Geophysical Research: Planets\/} {\bf 114} (\textit{Preprint} \eprint{https://agupubs.onlinelibrary.wiley.com/doi/pdf/10.1029/2008JE003294}) \urlprefix\url{https://agupubs.onlinelibrary.wiley.com/doi/abs/10.1029/2008JE003294}

\bibitem{PhysRevD.111.044061}
Majstorovi\ifmmode~\acute{c}\else \'{c}\fi{} J, Vidal L and Lognonn\'e P 2025 {\em Phys. Rev. D\/} {\bf 111}(4) 044061 \urlprefix\url{https://link.aps.org/doi/10.1103/PhysRevD.111.044061}

\bibitem{Stebbins1990}
Stebbins R~T and Bender P~L 1990 {\em AIP Conference Proceedings\/} {\bf 202} 188--204 ISSN 0094-243X (\textit{Preprint} \eprint{https://pubs.aip.org/aip/acp/article-pdf/202/1/188/11748825/188\_1\_online.pdf}) \urlprefix\url{https://doi.org/10.1063/1.39103}

\bibitem{StebbinsDrever1990}
Stebbins R~T, Armstrong J~W, Bender P~L, Drever R~W~P, Hellings R~W and Saulson P~R 1990 {\em AIP Conference Proceedings\/} {\bf 207} 637--646 \urlprefix\url{https://doi.org/10.1063/1.39356}

\bibitem{Izumi_2021}
Izumi K and Jani K 2021 {\em Detection Landscape in the deci-Hertz Gravitational-Wave Spectrum\/} (Springer Singapore) p 1–14 ISBN 9789811547027 \urlprefix\url{http://dx.doi.org/10.1007/978-981-15-4702-7_50-1}

\bibitem{Harms_2021}
Harms J, Ambrosino F, Angelini L, Braito V, Branchesi M, Brocato E, Cappellaro E, Coccia E, Coughlin M, Ceca R~D, Valle M~D, Dionisio C, Federico C, Formisano M, Frigeri A, Grado A, Izzo L, Marcelli A, Maselli A, Olivieri M, Pernechele C, Possenti A, Ronchini S, Serafinelli R, Severgnini P, Agostini M, Badaracco F, Bertolini A, Betti L, Civitani M~M, Collette C, Covino S, Dall'Osso S, D'Avanzo P, DeSalvo R, Giovanni M~D, Focardi M, Giunchi C, van Heijningen J, Khetan N, Melini D, Mitri G, Mow-Lowry C, Naponiello L, Noce V, Oganesyan G, Pace E, Paik H~J, Pajewski A, Palazzi E, Pallavicini M, Pareschi G, Pozzobon R, Sharma A, Spada G, Stanga R, Tagliaferri G and Votta R 2021 {\em The Astrophysical Journal\/} {\bf 910} 1 \urlprefix\url{https://dx.doi.org/10.3847/1538-4357/abe5a7}

\bibitem{Stavros}
{Katsanevas} S 2022 {Gravitational Wave Detection at the Moon} {\em 44th COSPAR Scientific Assembly. Held 16-24 July\/} vol~44 p 3042

\bibitem{Branchesi2023}
Branchesi M, Falanga M, Harms J, Jani K, Katsanevas S, Lognonné P, Badaracco F, Cacciapuoti L, Cappellaro E, Dell’Agnello S, de~Raucourt S, Frigeri A, Giardini D, Jennrich O, Kawamura T, Korol V, Landrø M, Majstorović J, Marmat P, Mazzali P, Muccino M, Patat F, Pian E, Piran T, Rosat S, Rowan S, St\"{a}hler S and Tissino J 2023 {\em Space Science Reviews\/} {\bf 219} ISSN 1572-9672 \urlprefix\url{http://dx.doi.org/10.1007/s11214-023-01015-4}

\bibitem{Lunarmodes}
Yan H, Chen X, Zhang J, Zhang F, Wang M and Shao L 2024 {\em Phys. Rev. D\/} {\bf 109}(6) 064092 \urlprefix\url{https://link.aps.org/doi/10.1103/PhysRevD.109.064092}

\bibitem{Creighton_et_al_inprep}
Creighton T {\em et~al.\/} 2025 (in preparation)

\bibitem{10.1117/12.3047830}
Diels J~C~M, Eikenberry S and Vanstryland E 2025 {Feasibility of a table-top gravitational wave detector} {\em Quantum Sensing, Imaging, and Precision Metrology III\/} vol PC13392 ed Shahriar S~M International Society for Optics and Photonics (SPIE) p PC133921W \urlprefix\url{https://doi.org/10.1117/12.3047830}

\bibitem{WATTERS2010}
Watters T~R, Robinson M~S, Beyer R~A, Banks M~E, Bell J~F, Pritchard M~E, Hiesinger H, van~der Bogert C~H, Thomas P~C, Turtle E~P and Williams N~R 2010 {\em Science\/} {\bf 329} 936--940 (\textit{Preprint} \eprint{https://www.science.org/doi/pdf/10.1126/science.1189590}) \urlprefix\url{https://www.science.org/doi/abs/10.1126/science.1189590}

\bibitem{Watters2025}
Watters T~R and Schmerr N~C 2025 {\em Science Advances\/} {\bf 11} eadu3201 (\textit{Preprint} \eprint{https://www.science.org/doi/pdf/10.1126/sciadv.adu3201}) \urlprefix\url{https://www.science.org/doi/abs/10.1126/sciadv.adu3201}

\bibitem{Horányi2015}
Horányi M, Szalay J, Kempf S, Schmidt J, Grün E and Srama R 2015 {\em Nature\/} {\bf 522} 324--326 \urlprefix\url{https://doi.org/10.1038/nature14479}

\bibitem{DEANGELIS2007169}
{De Angelis} G, Badavi F, Clem J, Blattnig S, Clowdsley M, Nealy J, Tripathi R and Wilson J 2007 {\em Nuclear Physics B - Proceedings Supplements\/} {\bf 166} 169--183 ISSN 0920-5632 proceedings of the Third International Conference on Particle and Fundamental Physics in Space \urlprefix\url{https://www.sciencedirect.com/science/article/pii/S0920563206010164}

\bibitem{WILLIAMS2017300}
Williams J~P, Paige D, Greenhagen B and Sefton-Nash E 2017 {\em Icarus\/} {\bf 283} 300--325 ISSN 0019-1035 lunar Reconnaissance Orbiter - Part II \urlprefix\url{https://www.sciencedirect.com/science/article/pii/S0019103516304869}

\bibitem{Glanzer_2023}
Glanzer J, Banagiri S, Coughlin S~B, Soni S, Zevin M, Berry C~P~L, Patane O, Bahaadini S, Rohani N, Crowston K, Kalogera V, Østerlund C, Trouille L and Katsaggelos A 2023 {\em Classical and Quantum Gravity\/} {\bf 40} 065004 \urlprefix\url{https://dx.doi.org/10.1088/1361-6382/acb633}

\end{thebibliography}

\end{document}